%% file: main.tex
\documentclass[aps, superscriptaddress, nofootinbib, preprintnumbers]{revtex4}

\include{header}

\begin{document}

\title{Effects of Critical Collapse on Primordial Black-Hole Mass Spectra}

\author{Florian K{\"u}hnel}
\email{florian.kuhnel@fysik.su.se}
\affiliation{The Oskar Klein Centre for Cosmoparticle Physics,
	Department of Physics,
	Stockholm University,
	AlbaNova,
	SE--106\.91 Stockholm,
	Sweden}
	
\author{Cornelius Rampf}
\email{cornelius.rampf@port.ac.uk}
\affiliation{Institute of Cosmology and Gravitation,
	University of Portsmouth,
	Dennis Sciama Building, Burnaby Road,
	Portsmouth PO1 3FX,
	United Kingdom}

\author{Marit Sandstad}
\email{marit.sandstad@astro.uio.no}
\affiliation{Nordita,
	KTH Royal Institute of Technology and Stockholm University,
	Roslagstullsbacken 23,
	SE--106\.91 Stockholm,
	Sweden}

\date{\formatdate{\day}{\month}{\year}, \currenttime}

\begin{abstract}
Certain inflationary models as well as realisations of phase transitions in the early Universe predict the formation of primordial black holes. For most mass ranges, the fraction of matter in the form of primordial black holes is limited by many different observations on various scales. Primordial black holes are assumed to be formed when overdensities that cross the horizon have Schwarzschild radii larger than the horizon. Traditionally it was therefore assumed that primordial black-hole masses were equal to the horizon mass at their time of formation. However, detailed calculations of their collapse show that primordial black holes formed at each point in time should rather form a spectrum of different masses, obeying critical scaling. Though this has been known for more than fifteen years, the effect of this scaling behaviour is largely ignored when considering predictions for primordial black hole mass spectra. In this paper we consider the critical collapse scaling for a variety of models which produce primordial black holes, and find that it generally leads to a shift, broadening and an overall decrease of the mass contained in primordial black holes. This effect is model and parameter dependent and cannot be contained by a constant rescaling of the spectrum; it can become important and should be taken into account when comparing to observational constraints.
\end{abstract}

\preprint{NORDITA-2015-131}

\maketitle

\section{Introduction}
\label{sec:Introduction}
\noindent
\input{Section--Introduction.tex}

\section{Critical Collapse}
\label{sec:Critical-Collapse}
\noindent
\input{Section--Critical-Collapse.tex}

\section{Models}
\label{sec:Models}
\noindent
\input{Section--Models.tex}

\subsection{Running-Mass Inflation}
\label{sec:Running--Mass-Inflation}
\noindent
\input{Section--Running--Mass-Inflation.tex}

\subsection{Hybrid Inflation}
\label{sec:Hybrid-Inflation}
\noindent
\input{Section--Hybrid-Inflation.tex}

\subsection{Axion-like Curvaton Inflation}
\label{sec:Axion--like-Inflation}
\noindent
\input{Section--Axion--like-Inflation.tex}

\subsection{First-Order Phase Transitions}
\label{sec:First--Order-Phase-Transitions}
\noindent
\input{Section--First--Order-Phase-Transitions.tex}

\section{Summary \& Outlook}
\label{sec:Summary-and-Outlook}
\noindent
\input{Section--Summary-and-Outlook.tex}

\acknowledgments

We thank Chris Byrnes, Bernard Carr, Maxim Khlopov, Peter Klimai, Dominik Schwarz, and Sam Young for useful discussions, and the anonymous Referee for his/her constructive comments that helped to improve the manuscript. F.K.~acknowledges supported from the Swedish Research Council (VR) through the Oskar Klein Centre. C.R.~acknowledges the support of the individual fellowship RA 2523/1-1 from the Deutsche Forschungsgemeinschaft.


\bibliography{refs}

\end{document}

%% file: header.tex
\usepackage{amsmath}	
\usepackage{amsthm}	
\usepackage{amssymb}	
\usepackage{datetime}
\usepackage{graphicx}
\usepackage{color}
\usepackage{verbatim}

\usepackage{hyperref}

\definecolor{grey}{rgb}{0.4,0.4,0.4}
\definecolor{dullmagenta}{rgb}{0.4,0,0.4}
\definecolor{darkblue}{rgb}{0,0,0.4}
\definecolor{midblue}{rgb}{0,0,0.5}
\definecolor{midred}{rgb}{0.5,0,0}
\definecolor{orange}{rgb}{1,0.5,0}
\definecolor{lightbrown}{rgb}{0.75,0.5,0.25}
\definecolor{tan}{cmyk}{0.14,0.42,0.56,0}
\definecolor{djunglegreen}{cmyk}{0.99,0,0.52,0}
\definecolor{lightgreen}{rgb}{0,1,0}
\definecolor{olivegreen}{cmyk}{0.64,0,0.95,0.40}
\definecolor{midgreen}{rgb}{0.0,0.675,0.0}
\definecolor{darkgreen}{rgb}{0,0.5,0}

\newcommand{\be}{\begin{equation}}
\newcommand{\ee}{\end{equation}}

\newcommand{\qq}{\qquad}
\newcommand{\qqq}{\qquad\quad}

\newcommand{\vs}{\vspace}

\renewcommand{\.}{\hspace{0.5mm}}

\newcommand{\ra}{\ensuremath{\rightarrow}}

\newcommand{\crm}{\ensuremath{\mathrm{c}}}

\newcommand{\erm}{\ensuremath{\mathrm{e}}}
\newcommand{\frm}{\ensuremath{\mathrm{f}}}
\newcommand{\grm}{\ensuremath{\mathrm{g}}}

\newcommand{\srm}{\ensuremath{\mathrm{s}}}

\newcommand{\Ocal}{\ensuremath{\mathcal{O}}}
\newcommand{\Pcal}{\ensuremath{\mathcal{P}}}

\renewcommand{\d}{\ensuremath{\mathrm{d}}}

\newcommand{\eg}{e.g.}
\newcommand{\ie}{i.e.}

\newcommand{\cf}{cf.} 

\let\baraccent=\= 
\renewcommand{\=}[1]{\stackrel{#1}{=}} 

\theoremstyle{definition}

\theoremstyle{remark}

\smallskipamount

\settimeformat{ampmtime}

\usepackage{float}



%% file: Section--Introduction.tex
Black holes could have formed in a primordial cosmological era from the gravitational collapse of density fluctuations $\delta = (\rho -\bar \rho)/\bar \rho$ on top of the mean energy density $\bar \rho$, shortly after a phase of accelerated expansion called inflation \cite{Brout:1977ix, Starobinsky:1980te, Kazanas:1980tx, Sato:1980yn, Guth:1982ec, Linde:1981mu, Linde:1983gd, Albrecht:1982wi} (for more recent reviews see \cite{Linde:2007fr, Martin:2013tda}). At such early times, most of the mean energy density of the Universe presumably stems from radiation. Formation of such primordial black holes (PBHs) \cite{ZeldovichNovikov69, Hawking:1971ei, Carr:1974nx} (see \cite{Carr:2009jm, Khlopov:2008qy} for a recent reviews) could arise when a critical threshold $\delta_{\crm}$ is exceeded such that the radiation pressure, acting as a counter force to gravity, cannot prevent the collapse anymore. 

Primordial black holes, if existent, could have observational implications for current and ongoing surveys, for example through their gravitational interactions or because of their evaporation via Hawking radiation ($T_{\rm Hawking} \sim 1 / M$) \cite{Hawking:1974rv, Hawking:1974sw}. In certain mass ranges, primordial black holes could even make up (parts of) the not yet identified cold dark matter \cite{Frampton:2010sw}, but constraints arising from the standard model of cosmology limit this possibility drastically \cite{Bergstrom:2000pn,Carr:2009jm}. For example, constraints from the Big Bang nucleosynthesis, arising from entropy limits --- or limits on the abundance in light elements, constrain the abundant primordial black-hole masses in the range between $10^6 -10^{13}$g \cite{Suyama:2004mz}. Primordial black holes with masses less than $10^{15}$\,g would have already evaporated by the present time, due to Hawking radiation, and the effects of their evaporation might have been observable had they existed in sufficient abundance. Specifically, if primordial black holes with a mass of about $10^{15}$\,g had existed in sufficient abundance, we would measure an excess of photons with energy of about $100$\,MeV today, which is not observed in the $\gamma$-ray background. The non-detection of such black holes implies that their density has to be smaller than $10^{-8}$ times the critical density \cite{Carr:2009jm}. There are various other mechanisms which observationally constrain the mass range of primordial black holes. An overview of observational limits can be found in Ref.~\cite{Carr:2009jm} (\cf~Fig.~9 therein). For more recent constraints that stem from the capture of primordial black holes by white dwarfs and neutron stars see Refs.~\cite{Capela:2012jz, Capela:2013yf, Capela:2014ita}.

All production mechanisms for primordial black holes have one essential feature in common, that is the appearance of \textit{critical phenomena}, a fact established through state-of-the-art general relativistic numerical computations \cite{Choptuik:1992jv,Niemeyer:1998ac, Musco:2004ak, Musco:2008hv, Musco:2012au}. When sufficiently large fluctuations (re-)enter the particle horizon ($r_{H} \sim t$) and collapse to a black hole, one could na{\"i}vely expect that the black hole mass should be of the same order as the mass associated with the current horizon patch; this associated horizon mass is roughly $M_{H} \simeq 5 \times 10^{4}\.M_{\odot}\,t [ \srm ]$ (\cf~\cite{Carr:2009jm}). Most of such large fluctuations will be close to the critical threshold $\delta_{\crm}$, because even larger fluctuations are in the tail of the (almost) Gaussian distribution of primordial density fluctuations, and thus suppressed. General relativistic numerical computations performed in the literature modeled the collapse of a relativistic (and for some cases specifically a perfect) fluid, which is supposably the continuum description of the considered primordial black-hole formation. These results indicated that the resulting black hole mass distribution function does not peak at the horizon mass, but instead exhibits a \textit{spectrum of black hole masses}. Consider any one-parameter family $p$ (\eg, the density contrast) of regular asymptotically flat initial data such that the space-time becomes flat if $p < p_{\crm}$ and forms a black hole for $p > p_{\crm}$. Then, near the critical threshold $p_{\crm}$, the mass spectrum of black holes obeys the following scaling law,
\begin{align}
	M( p )
		&=
								k\.M_{H}
								\left(
									p
									-
									p_{\crm}
								\right)^{\gamma}
								\, ,
								\label{eq:scaling}
\end{align}
where $M_{H}$ is the associated horizon mass, $k$ is a dimensionless constant and the critical exponent $\gamma$ is universal with respect to the initial data (for a recent review see \eg~\cite{Gundlach:2007gc}). It is evident from this scaling law that black holes with arbitrarily small masses can be created.\footnote{Note in particular the case $M \to 0$ when $p \to p_{\crm}$ which is fundamentally different from the approximation of direct horizon-mass collapse.}

When the critical-scaling behaviour in the formation of primordial black holes was first discovered and explored \cite{Choptuik:1992jv, Niemeyer:1998ac}, its implications on mass distributions near monochromatic spectra was considered and compared to broader mass spectra with no critical collapse \cite{Yokoyama:1998xd, Green:1999xm}. At the time, highly peaked spectra for the primordial black holes was what was mostly considered and hence the effect of critical collapse scaling was thought to be roughly degenerate with more realistic initial mass spectra. In the literature, therefore, with notable exceptions such as \cite{Bugaev:2008gw}, one mostly approximates the primordial black-hole mass to be identical with the horizon mass{\,---\,}or one simply scales the overall spectrum by some constant. Both approaches yield a specific functional form for the predictions of $\beta$ in any given model. Crucially, as we shall show in this work, using instead a spectrum of primordial black-hole masses consistent with the critical collapse as given by Eq.\,\eqref{eq:scaling}, the resulting fractional density $\beta$ gets \emph{shifted}. Neglecting that shift of $\beta$ could change the observational consequences tremendously for certain models predicting primordial black holes, since some constraints are very sharply located in mass space (\cf~Fig.~9 in Ref.~\cite{Carr:2009jm}). Hence in the most extreme scenarios, models could be ruled out or favoured again because of that shift. For instance if a model predicts a high abundance around a mass range where some masses are highly constrained and there is a sharp transition to a much less constrained region, the shift of the spectrum towards lower masses due to critical collapse may move the entire spectrum of primordial black holes from an unconstrained region to a constrained region or vice versa.

Though most standard models for inflation do not predict primordial black-hole production, several viable models of inflation still predict a possible abundance in primordial black holes. In this paper we will study the effect of the critical collapse scaling \eqref{eq:scaling} on the mass distribution functions of the primordial black holes of three such models: running-mass inflation \cite{Drees:2011hb}, hybrid inflation \cite{Clesse:2015wea}, and axion-like curvaton inflation \cite{Kawasaki:2012wr}. Apart from some inflationary models, there are other mechanisms in the early Universe which could produce primordial black holes, amongst them are phase transitions (\eg, \cite{Jedamzik:1999am}) for which we shall also investigate how the mass distribtution function of primordial black holes is affected. Specifically, first-order phase transitions are usually accompanied with a change in pressure (or more precisely the speed of sound with which pressure is mediated), which could imply that the threshold $\delta_{\crm}$ for primordial black-hole formation is (suddenly) reduced when the pressure decreases. Alternatively, formation due to phase transitions could also arise without any prior inhomogeneities, for example from so-called bubble collisions, domain walls or cosmic strings (see \cite{Carr:2009jm} and references therein), but in this paper we shall not investigate such alternative scenarios. 

This paper is organised as follows. In Sec.~\ref{sec:Critical-Collapse} we will introduce the critical collapse of primordial black holes in some greater detail, focusing mainly on what we have used to account for this in the different models. In Sec.~\ref{sec:Models} we will briefly review the specific models and study the effects of critical collapse to the produced mass spectra for primordial black holes in each of them. Finally, we give a Summary and Outlook in Sec.~\ref{sec:Summary-and-Outlook}.

%% file: Section--Critical-Collapse.tex
While it might appear reasonable to assume that once sufficiently large overdensities re-enter the horizon they immediately collapse to a black hole of horizon mass $M_{H}$, a more refined treatment of the collapse exhibits a so-called {\it critical scaling} spectrum for the primordial black-hole mass distribution $M$ \cite{Choptuik:1992jv, Koike:1995jm,Gundlach:1999cu,Gundlach:2002sx} of the form [\cf~Eq.~\eqref{eq:scaling}]
\begin{align}
	M
		&=
								k\.M_{H}\.
								\big(
									\delta
									-
									\delta_{\crm}
								\big)^{\gamma}
								\; ,
								\label{eq:M-delta-scaling}
\end{align}
where $\delta > \delta_{\crm}$. The constant $k$, the threshold $\delta_{\crm}$ as well as the critical exponent $\gamma$ depend on the specific fluid the overdensity $\delta$ is re-entering into \cite{Musco:2012au}. Besides the mentioned theoretical considerations on the critical collapse, there have been in-depth numerical confirmations \cite{Niemeyer:1999ak, Musco:2004ak, Musco:2008hv, Musco:2012au} (\cf~in particular Fig.~1 of Ref.~\cite{Musco:2008hv} for justifying the above scaling law).

Right after the discovery of the critical collapse of primordial black holes in the 1990s \cite{Choptuik:1992jv,Niemeyer:1998ac}, this effect was considered \cite{Yokoyama:1998xd,Green:1999xm}, and the conclusion was that the horizon-mass approximation was still reasonably good. Despite the fact that now more precise observational limits are available, using this approximation seems to be the general approach taken by the field (\cf~the recent reviews \cite{Green:2014faa, Carr:2009jm}).

The topic of this paper is to reinvestigate the critical collapse, particularly in light of current observational constraints on primordial black holes. Since these constraints are partly very stringent in certain regions of the mass spectrum \cite{Carr:2009jm}, the changes to the mass distribution that we obtain in the subsequent section (Sec.~\ref{sec:Models}) should not be ignored. Instead{\,---\,}as we shall demonstrate in the present work{\,---\,}the continued use of the horizon-mass approximation turns out to be a source of potentially very large errors.

In general, the critical exponent $\gamma$ seems to be independent of the concrete perturbation profile \cite{Neilsen:1998qc, Musco:2012au}, though $\delta_{\rm c}$ and $k$ may depend on this. Throughout this work we shall apply the Press--Schechter formalism \cite{1974ApJ...187..425P} on spherical collapse (\eg~\cite{Green:2004wb}), using a Gaussian perturbation profile, \ie
\begin{align}
	\Pcal( \delta )
		&\equiv
								\frac{ 1 }{ \sqrt{2 \pi \sigma^{2}\,} }\,
								\exp\!
								\left(
									- \frac{ \delta^{2} }{ 2\.\sigma^{2} }
								\right)
								,
								\label{eq:P-Gaussian}
\end{align}
which is in good accordance with current measurements of the cosmic microwave background (CMB) \cite{Ade:2015ava}. The quantity $\sigma$ is the variance of the primordial power spectrum of density perturbations coming from the appropriate model of inflation.

In radiation domination the value of $\gamma$ has been found in repeated studies to be about $\gamma \simeq 0.36$ \cite{Koike:1995jm, Niemeyer:1999ak, Musco:2004ak, Musco:2008hv, Musco:2012au}. A good approximation for $\delta_{\rm c}$ has been found to lie in the regime $0.41-0.45$ \cite{Musco:2004ak, Musco:2008hv, Musco:2012au}.\footnote{Note that these results rely on the assumption of spherical collapse which was proved to be a good assumption in \cite{Doroshkevich1970,1986ApJ...304...15B}. For non-spherical collapse the results might differ significantly (\cf~\cite{Sheth:1999su}). We leave a respective investigation for future studies.} 
We have used the value of $0.45$ where possible, however, we have also chosen other values where appropriate; either for physical reasons in the section on phase transitions (Sec.~\ref{sec:First--Order-Phase-Transitions}) or to allow for comparison with the approach taken in the literature in the case of hybrid inflation (Sec.~\ref{sec:Hybrid-Inflation}). In accordance with \cite{Niemeyer:1997mt} we also chose to set $k = 3.3$. In a realistic treatment we expect that this value for $k$ might not be the most accurate choice, however, we expect that the general tendencies that we find will be the same. The exact values used for $\gamma$ and $\delta_{\rm c}$ will be specified for each case.

A convenient measure of how many primordial black holes are being produced can be given by the ratio of the energy density in primordial black holes by the total energy density
\begin{align}
	\beta
		&\equiv
								\frac{ \rho_{\rm PBH} }
								{\rho_{\rm tot}}
								\; .
								\label{eq:beta-definition}
\end{align}
Within the Press--Schechter formalism applied to black-hole formation, one can express $\beta$ as \cite{Niemeyer:1997mt}
\begin{align}
	\beta
		&=
								\int_{\delta_{\crm}}^{\infty}
								\d \delta\;
								k
								\big(
									\delta
									-
									\delta_{\crm}
								\big)^{\!\gamma}_{}\,
								\Pcal( \delta )
		\approx
								k\.\sigma^{2\gamma}\,
								{\rm erfc}
								\bigg(
									\frac{ \delta_{\crm} }{ \sqrt{2\,}\.\sigma }
								\bigg)
								\; ,
								\label{eq:Beta_normalisation} 
\end{align}
where we used $\sigma \ll \delta_{\crm}$. We have numerically confirmed the validity of this approximation for our considerations. In the first line we have extended the upper integration above $\delta = 1$, in contrast to what was done in \cite{Niemeyer:1997mt}. This has been shown in Ref.~\cite{Kopp:2010sh} not to lead to a separate-Universe production. Therein the authors point out that the previous choice of this limit is in fact gauge dependent. This topic is still subject to discussion \cite{Carr:2014pga}, in practice, however, the integrand for large values of $\delta$ is so small that the results of taking one choice or the other are nearly equivalent.
Also note that alternatively to using the Press--Schechter formalism, one could calculate the mass fraction $\beta$ in terms the so-called peaks-theory approach \cite{Bardeen:1985tr}. There is some recent discussion (\cf~\cite{Young:2014ana}) on whether the mentioned formalisms predict differences in the overall amplitude of $\beta$. However, the effect of critical collapse also concerns broadening and shift towards smaller masses{\,---\,}signatures which are distinctive for the critical scaling and should hold irrespective of the use of either the mentioned formalisms. The same holds true for the extension of the Press--Schechter formalism to solve the so-called cloud-in-cloud problem (\cf~\cite{Jedamzik:1994nr}).

Following \cite{Niemeyer:1997mt} we next derive the primordial black-hole initial mass function $g$, defined as the primordial black-hole number $\d n_{\rm PBH}$ per mass interval $\d M$,\footnote{Note the slight difference in the definition of the initial mass function as compared to the one in Ref.~\cite{Niemeyer:1997mt}.} 
\begin{align}
	g
		&\equiv
								\frac{ \d n_{\rm PBH} }{ \d M }
		\approx
								\frac{m^{\frac{ 1 }{ \gamma } - 1}\,
 									\exp{\!
										\left[
											-
												\left(
													\delta
													+
													m^{\frac{ 1 }{ \gamma }}
													\right)^{2} /
												\,\big( 2 \sigma ^{2} \big)
										\right]
										}
									}
									{ \sqrt{2 \pi\,}\.\gamma\.\sigma\,
										{\rm erfc}\!
										\left(
											\frac{\delta}
											{\sqrt{2\,} \sigma }
										\right)
								}
								\; ,
								\label{eq:PBH-IMF-definition}
\end{align}
which again holds for $\sigma \ll \delta_{\crm}$, and we have defined $m \equiv M / ( k\.M_{H} )$. In deriving Eq.~\eqref{eq:PBH-IMF-definition} we have made use of the Gaussian profile \eqref{eq:P-Gaussian} for the amplitude of the fluctuations.

In the subsequent sections (\ref{sec:Running--Mass-Inflation}-D) we apply this critical collapse to a representative set of models for production of primordial black holes. In practice the procedure we have employed to account for this is the following: We have taken the perturbation spectrum of primordial density fluctuations and binned it. Each bin then corresponds to a particular horizon mass $M_{H}$ and a formation time at which the horizon was of the corresponding size. For each model and parameter set, this bin has a particular value of the variance $\sigma$. For each bin we have then calculated the initial mass function $g$ given in Eq.~\eqref{eq:PBH-IMF-definition}. We have used this initial mass function and normalised it according to Eq.~\eqref{eq:Beta_normalisation} to get the spectrum $\beta$ at formation time according to critical collapse by multiplying $g$ by the value of the mass and the mass interval. We then take into account the time evolution of the primordial black-hole (matter) density with respect to the background (radiation) energy density until matter-radiation equality.\footnote{We assume that matter-radiation equality occurs at about 47,000 years after the Big Bang, in accordance with the currently favoured $\Lambda$CDM model \cite{Ade:2015lrj}.} In the case of $\rho_{\rm PBH} \ll \rho_{\rm tot}$, $\beta$ grows approximately linearly with growing scale factor. The full spectrum for the considered model is then obtained by adding the spectra for all the bins into one function. It is important to note that, here, the binning process in itself is purely a numerical tool and has no practical consequences. Though the size of the bins is used in the multiplication by the initial mass function, the effect of the particular binning is eliminated by the normalisation of it. Since what we are binning is a continuous function, we can increase the number of bins indefinitely. In practice we have increased the number of bins until we have reached convergence of the resulting mass function, so the result is in practice equal to what it would have been for a procedure done at each point.  This also implies that there won't be any problem in counting isolated overdense regions (\cf~the cloud-in-cloud problem \cite{Jedamzik:1994nr}). Finally, we compare the results of $\beta$ to those obtained by evolving a horizon mass collapse spectrum from formation to matter-radiation equality.

%% file: Section--Models.tex
In this section we elaborate on the importance of the inclusion of critical collapse for various important models of primordial black-hole production. Specifically, we will show the consequences of the associated shift and broadening of the produced spectra of primordial black holes in {\it running-mass inflation} (Sec.~\ref{sec:Running--Mass-Inflation}), {\it hybrid inflation} (Sec.~\ref{sec:Hybrid-Inflation}), {\it axion-like curvaton inflation} (Sec.~\ref{sec:Axion--like-Inflation}), and {\it first-order phase transitions} (Sec.~\ref{sec:First--Order-Phase-Transitions}). These models represent along with (p)reheating and other types of coupling to particle production, the main sources of viable production of primordial black holes known at present \cite{Clesse:2015wea}.
 
We stress that the aim of this article is \emph{not} to investigate the most realistic parameter sets for the individual models. Rather, we demonstrate the importance of critical scaling for any model. We will show that neglecting this scaling can in certain cases result in relative errors far larger than $100\.\%$ with respect to the shape of the specific primordial black-hole abundance, \eg~in the location of its peak and height. Though the model parameters that we use are not necessarily the most realistic, we have chosen parameters comparable to or in ranges mentioned to be relevant for primordial black hole production in the various models in their considerations in the literature. Hence though the parameters may not be realistic, for the most part they will be as realistic as the parameters considered in the literature. 
We also note that the parameter space of multifield models of inflation 
could be constrained further by the bound on primordial non-Gaussianity ($| f_{\rm nl}^{\rm local} | \lesssim 0.001$) \cite{Young:2015kda}. The presence of the latter could have a significant effect on the abundance of primordial black holes and potentially overproduce isocurvature modes in the CMB which are constrained by current observations \cite{Ade:2015lrj}. For related discussions we refer to Refs.~\cite{Bugaev:2011qt, Bugaev:2011wy, Young:2015kda, Tada:2015noa}.

The key ingredient for any model to produce primordial black holes is that it needs to generate curvature perturbations larger than some threshold value $\delta_{\crm}$ at a given early time which then collapse into black holes after horizon re-entry. As the amplitude of the curvature-perturbation power spectrum at the pivot scale $k_{\star} = 0.002\.{\rm Mpc}^{-1}$ for the CMB measurements of WMAP \cite{Komatsu:2010fb} or $k_{\star} = 0.005\.{\rm Mpc}^{-1}$ for Planck \cite{Planck:2013jfk,Ade:2015lrj} is far too small to produce a notable abundance of primordial black holes, one needs the power spectrum to become large at an early time. For the case of a power spectrum which monotonically increases with decreasing $k$, the primordial black-hole abundance is largest for smallest $M_{H}$. The exact value of the threshold $\delta_{\crm}$ depends on the precise medium the perturbations re-enter into (\cf~the discussion in the previous section, and also \cite{Musco:2008hv}).

The models we are discussing in the following subsections all have in common that their power spectra have an increased amplitude feature, \ie~a bump or spike, at some small scale/high $k$, with the details depending on the specific model parameters. For a moderate number (three or four) of representative parameter sets, we subsequently investigate the effect of critical collapse and show how each of the spectra change.

Before going through the specifics of the mentioned models, let us briefly comment on the primordial black-hole production in preheating \cite{Green:2000he}. Here, the idea is that black holes are being produced during the inflaton decay at the end of inflation. However, as the horizon mass at that time is rather small, the associated black holes will also be, and so more or less decay right after their production, which prevents them from constituting a viable dark-matter contribution today. Their main effect will be additional heating. A large overproduction might still be in conflict with observations, and the study of primordial black hole production in preheating can thus lead to constraints on preheating models \cite{Torres-Lomas:2014bua}. Now, as we expect that critical collapse will broaden and shift the mass distribution, the mentioned additional heating will be less effective as compared to the standard case. We leave a corresponding investigation to future work. 

Primordial black-hole formation can also be triggered by interactions between the inflaton and other fields, leading to the production of non-inflaton particles during inflation \cite{Lin:2012gs, Linde:2012bt, Bugaev:2013fya}. This can then even lead to a production of primordial black holes which might still be present in the Universe today \cite{Erfani:2015rqv}. We will not consider the critical-collapse treatment of the primordial black-hole production from these sources here, but defer their consideration to future works. However, on general grounds we will argue that bounds obtained from non-production of primordial black holes in these models may in practice change and will presumably be less stringent, when critical collapse is taken into account.

We note that the recently proposed corpuscular description of black holes on the full quantum level via Bose--Einstein condensates of gravitons \cite{Dvali:2012en, Dvali:2011aa} might lead to strong constraints on the production period/mass of primordial black holes: This framework naturally predicts baryon-number conservation \cite{Dvali:2012rt, Kuhnel:2015qaa}, which leads to a bound on primordial black-hole production with mass below approximately $10^{-7}\.M_{\odot}$ \cite{Kuhnel:2015qaa}. However, as it is yet to be understood what kind of objects are produced at smaller masses instead, and how they possibly constitute dark matter, we will focus on investigating how the critical collapse alters the classical results on the primordial black-hole production, and leave the mentioned corpuscular studies for future work.

%% file: Section--Running--Mass-Inflation.tex
\vs{-6mm}

Primordial black-hole formation in the running-mass model \cite{Stewart:1996ey, Stewart:1997wg} has been intensively studied in amongst other works \cite{Drees:2011yz, Drees:2011hb, Drees:2012sz} (\cf~also~\cite{Leach:2000ea} for a discussion on constraints).\footnote{In \cite{Bugaev:2008gw}, critical-collapse effects were also studied in a particular version of running mass inflation, and noted to be considerable. The version we study here is slightly different as it follows a very simple expansion order for order in the spectral index which is easier to compare both to observational bounds \cite{Komatsu:2010fb,Planck:2013jfk,Ade:2015lrj} and stays close to what was done in \cite{Drees:2011hb}.} The perhaps simplest realisation may be expressed through the inflationary potential\vs{-2mm}
\begin{align} \label{pot1}
	V( \phi )
		&=
								V_{0}
								+
								\frac{ 1 }{ 2 }\.m_{\phi}^{2}(\phi) \phi^{2}
								\, ,
\end{align}
with the constant $V_{0}$, and the scalar field $\phi$.

There exists a plethora of embeddings of this model in various frameworks such as hybrid inflation \cite{Linde:1993cn} for instance, which lead to different specific functions $m_{\phi}( \phi )$. These yield distinct expressions for the primordial density power spectra whose variance can be recast into the general form \cite{Drees:2011hb}:
\begin{align}
	\big[ \sigma( k ) \big]^{2}
		&\simeq
								\frac{ 8 }{ 81 }\.\Pcal( k_{\star} )
								\bigg(
									\frac{ k }{ k_{\star} }
								\bigg)^{\!\! n( k ) - 1}
								\Gamma\!
								\left(
									\frac{ n_{\srm}( k ) + 3 }{ 2 }
								\right)
								\label{eq:sigma-running-mass}
								,
\end{align}
where the spectral indices $n( k )$ and $n_{\srm}( k )$ are given by 
\begin{subequations}
\begin{align}
	n( k )
		&=
								n_{\srm}( k_{\star} )
								-
								\frac{ 1 }{ 2! }\.a\.\ln\!
								\bigg(
									\frac{ k }{ k_{\star} }
								\bigg)
								+
								\frac{ 1 }{ 3! }\.b\.\ln^{2}\!
								\bigg(
									\frac{ k }{ k_{\star} }
								\bigg)
								-
								\frac{ 1 }{ 4! }\.c\.\ln^{3}\!
								\bigg(
									\frac{ k }{ k_{\star} }
								\bigg)
								+
								\ldots
								\; ,
								\label{eq:n-running-mass}
								\displaybreak[1]
								\\[2mm]
	n_{\srm}( k )
		&=
								n_{\srm}( k_{\star} )
								-
								a\.\ln\!
								\bigg(
									\frac{ k }{ k_{\star} }
								\bigg)
								+
								\frac{1}{2}\.b\.\ln^{2}\!
								\bigg(
									\frac{ k }{ k_{\star} }
								\bigg)
								-
								\frac{1}{6}\.c\.\ln^{3}\!
								\bigg(
									\frac{ k }{ k_{\star} }
								\bigg)
								+
								\ldots
								\; ,
\end{align}
\end{subequations}
with real parameters $a$, $b$, and $c$. The terms multiplied by $a$ are referred to as ``running'' terms, those multiplied by $b$ are called ``running-of-running'', and the $c$ terms are dubbed ``running-of-running-of-running''. The expansion is in principle infinite to account for any functional form of the running at any value of $k$, however here we chose to consider a model that includes only the first few terms so as to compare with what has been done in \cite{Drees:2011hb} extending their analysis with one extra order as will be explained below.

As the spectral index and amplitude of the primordial power spectrum at the pivot scale $k_{\star} = 0.002\.{\rm Mpc}^{-1}$ have been measured to be $n_{\rm s}( k_{\star} ) \approx 0.96 < 1$ and $\Pcal( k_{\star} ) = \Ocal( 10^{-9} )$, respectively \cite{Komatsu:2010fb,Planck:2013jfk,Ade:2015lrj}, models without running certainly cannot produce primordial black holes in any notable abundance. Furthermore, with the measurement of $a = −0.003 \pm 0.007 \ll 1$ \cite{Ade:2015lrj}, running alone is not enough to give a sufficient increase of the power spectrum at early times. Hence one needs at least to include a running-of-running term (which is only weakly constrained: $b \simeq 0.02 \pm 0.02$ \cite{Planck:2013jfk,Ade:2015lrj}).

\begin{figure}
	\centering
	\includegraphics[scale=1,angle=0]{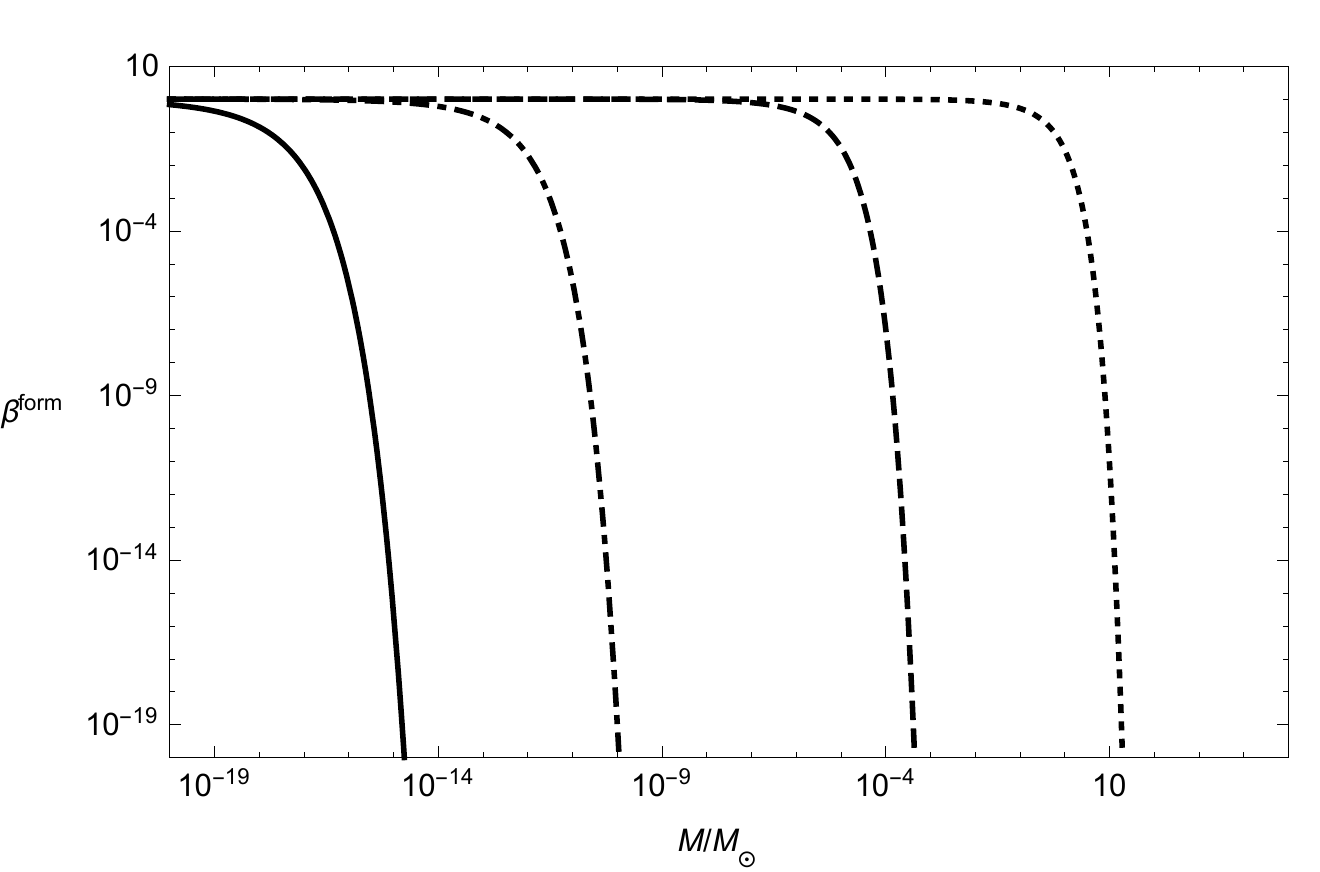}
	\caption{The relative energy density $\beta^{\rm form}$ for the running-mass case 
		including running, and running-of-running	due to Eq.~\eqref{eq:sigma-running-mass}, 
		at the time of formation as a function of $M / M_{\odot}$.
		For all graphs we set $a = 0.011$; individually we use
		$b = 0.0050$ (dotted), 
		$b = 0.0025$ (dashed), 
		$b = 0.0010$ (dot-dashed), as well as 
		$b = 0.0005$ (solid).
		For all curves the threshold $\delta_{\crm}$ has been set to $0.45$.
		Note that {\it all} curves increase towards lower masses, 
		yielding inevitable overproduction of small primordial black holes.}
	\label{fig:beta-form-without-running^3}
\end{figure}

Figure~\ref{fig:beta-form-without-running^3} shows the ratio $\beta$ for four of these cases (\cf~its figure caption for information about the respective parameters). In this Figure, exceptionally, we do {\it not} plot $\beta$ at the time of radiation-matter equality (as we do in all the other cases), but rather at the time of primordial black-hole formation. The reason being that, although basically any desired abundance of primordial black holes {\it of a specific mass} can be generated, due to the monotonic nature of $\beta$, there will {\it always} be an enormous over-production of small black holes. This {\it inevitably} violates respective bounds on their abundance \cite{Carr:2009jm} and there is no need to time-evolve the results to equality. In the treatment of the topic in \cite{Drees:2011hb}, this overproduction at small scales is not considered as their aim is to show that sufficiently many large primordial black holes can be produced to account for the dark matter present in our Universe. However, the bounds on these very small primordial black holes are very severe \cite{Carr:2009jm}, and should not be neglected, especially as this can be done within the same scheme as we will show below.

\begin{figure}
	\centering
	\includegraphics[scale=1,angle=0]{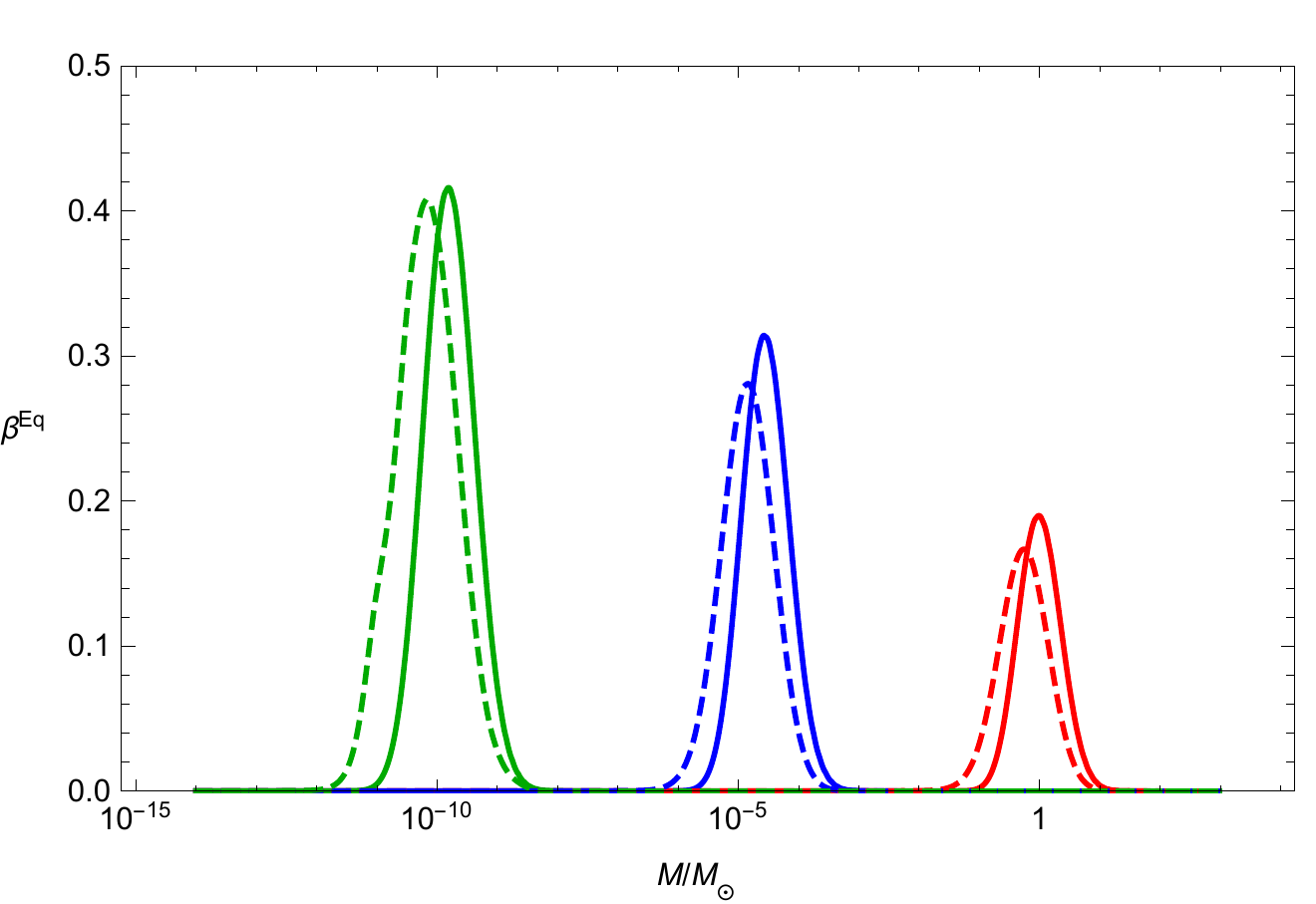}
	\caption{The relative energy density $\beta^{\rm Eq}$ for the running-mass case 
		including running, running-of-running, and running-of-running-of-running, 
		due to Eq.~\eqref{eq:sigma-running-mass}, at the time of radiation-matter equality 
		as a function of $M / M_{\odot}$.
		The solid curves assume standard black-hole production with horizon mass, 
		where $a = 0.011$ for all graphs, and individually (right to left) 
		$b = 0.02$, $c = - 0.00234$ (red), 
		$b = 0.011$, $c = - 0.0010975$ (blue), 
		$b = 0.006$, $c = - 0.000519$ (green). 
		For all curves the threshold $\delta_{\crm}$ has been set to $0.45$. 
		The dashed curves have the same respective parameters, 
		but assume critical collapse according to Eq.~\eqref{eq:M-delta-scaling} with $\gamma = 0.36$.}
	\label{fig:beta-Eq-Running-Mass-Comparison}
\end{figure}

In order for the power spectrum (and hence $\beta$) to prevent this overproduction of small primordial black holes, one needs to include higher orders (than the second order) in the expansions (\ref{eq:n-running-mass},b). Figure~\ref{fig:beta-Eq-Running-Mass-Comparison} shows the effect of the inclusion of running-of-running-of-running on $\beta$ at the time of matter-radiation equality (``Eq''). The dashed lines depict the influence of critical collapse. As expected, we observe a shift towards lower masses as well as a broadening with respect to the horizon-mass case. The associated areas are approximately cut in half.

%% file: Section--Hybrid-Inflation.tex
Hybrid inflation is a two-field inflationary framework, which was first introduced by Andrei Linde in 1993 \cite{Linde:1993cn}. It generically describes the situation in which inflation ends in a rapid rolling (``waterfall'') phase of one scalar field, which is triggered by the other one.

When sufficiently large curvature perturbations, which produce primordial black holes (in standard hybrid-inflation), re-enter shortly after the end of inflation, their masses are relatively low, \ie~$\lesssim \Ocal(10^{-29}\.M_{\odot})$, such that they cannot constitute a significant fraction of dark matter today. Here we re-investigate a recently proposed scenario by Clesse and Garc{\'\i}a-Bellido \cite{Clesse:2015wea} using similar parameters. Therein, inflation ends with a mild waterfall phase which typically lasts for dozens of $e$-folds, yielding a possible production of primordial black holes on mass ranges up to about $\Ocal(10^{2}\.M_{\odot})$, where the width of the associated spectra can span several orders of magnitude, depending on the specific model parameters.

We introduce the mentioned model, following \cite{Clesse:2015wea}, and specify the two-field potential:
\begin{align}
\label{eq:potential}
	V( \phi, \psi )
		&=
								\Lambda
								\left[
									\left(
										1
										-
										\frac{ \psi^{2} }
										{ M^{2} }
									\right)^{2}
									+
									\frac{ \phi - \phi_{\crm} }
									{ \mu_{1} }
									- \frac{ ( \phi - \phi_{\crm} )^{2} }
									{ \mu_{2}^{2} }
									+
									\frac{ 2 \phi^{2} \psi^{2} }
									{ M^{2} \phi_{\crm}^{2} }
								\right]
								,
\end{align}
with $\mu_1 \simeq 3 \times 10^5\.M_{\rm Pl}$ and $\mu_{2} \simeq 10\.M_{\rm Pl}$, leading to
\begin{align}
	\Pcal_{\zeta}( k_{\star} )
		&\simeq
								\frac{ \Lambda\.\mu_{1}^{2} }
								{ 12 \pi^{2}\.M_{\rm Pl}^{6} }
								\left(
									\frac{ k_{\star} }{ k_{\phi_{\crm}} }
								\right)^{\!n_{\srm} - 1}
		=
								2.21 \times 10^{-9}
								\; ,
\end{align}
\if\case
which relates the $\Lambda$ and $\mu_{1}$ parameters via
\begin{align}
	\Lambda
		&=
								2.21 \times 10^{-9}
								\times
								\frac{ 12\.\pi^{2}\.M_{\rm Pl}^{6}}
								{ \mu_{1}^{2} }
								\left(
									\frac{ k_{\phi_{\crm}} }
									{ k_{\star} }
								\right)^{n_{\srm} - 1}
								,
								\label{eq:Lambda}
\end{align}
\fi
where the ratio $k_{\phi_{\crm}} / k_{\star}$ is determined assuming instantaneous reheating and solving the two-field dynamics such that the scalar power spectrum amplitude matches the value derived from Planck \cite{Planck:2013jfk}.

Using the above, the variance $\sigma$ of curvature perturbations can be related to the power spectrum through \cite{Clesse:2015wea}
\begin{align}
	\sigma( k )
		&\simeq
	\sqrt{\Pcal_{\zeta}( k )\,}
		\simeq
								\sqrt{	\frac{ \Lambda\.M^{2}\.\mu_{1}\.\phi_{\crm} }
								{ 192\.\pi^{2}\.M_{\rm Pl}^{6}\.\kappa_{2}\.\psi_{k}^{2} }\,}
								\; .
								\label{eq:sigma-hybrid}
\end{align}
Here, $\psi_{k} \equiv \psi_{0}\.\erm^{\kappa_{k}}$, $\kappa_{k} \equiv 4\.\phi_{\crm}\.\mu_{1} \xi_{k}^{2}\./\.M^{2}$ and $\xi_{k} \equiv - M_{\rm Pl}^{2} \.( N_{1} + N_{2} - N_{k} ) / ( \mu_{1} \phi_{\crm} )$, with
\begin{align}
	\psi_{0}
		&\equiv
								\left(
									\frac{ \Lambda\.\sqrt{2\.\phi_{\crm}\.\mu_{1}^{}\,}\.M }
									{ 96\.\pi^{3/2} }
								\right)^{1/2}
								\label{eq:psi0}
\end{align}
as well as
\begin{align}
	\xi_{2}
		\equiv
								-
								\frac{ M \sqrt{\kappa_{2}\,} }{ 2\.\sqrt{\mu_{1}\.\phi_{\crm}\,} }
								\; ,
\end{align}
and the durations of the two phases are given, in terms of numbers of $e$-foldings, by
\begin{align}
	N_{1}
		&=
								\frac{ \sqrt{\kappa_{2}\.\phi_{\crm}\.\mu_{1}\,} M }
								{ 2 M_{\rm Pl}^{2} }
								\; ,\qq
	N_{2}
		\simeq
								\frac{ M \sqrt{ \mu_{1}\.\phi_{\crm}\,} }
								{ 4\.M_{\rm Pl}^{2}\.\sqrt{\kappa_{2}\,} }
								\; .
\end{align}
For a given wavenumber $k$, exiting the Hubble radius $| N_{k} |$ $e$-foldings before the end of inflation, the associated primordial black-hole mass when assuming horizon-mass collapse is given by \cite{Clesse:2015wea}
\begin{align}
	M_{k}
		&=
								\frac{ M_{\rm Pl} }{ \sqrt{ \Lambda / 3\,} } \erm^{- 2 N_{k}}
								\; .
								\label{eq:Mk}
\end{align}

\begin{figure}
	\centering
	\includegraphics[scale=1,angle=0]{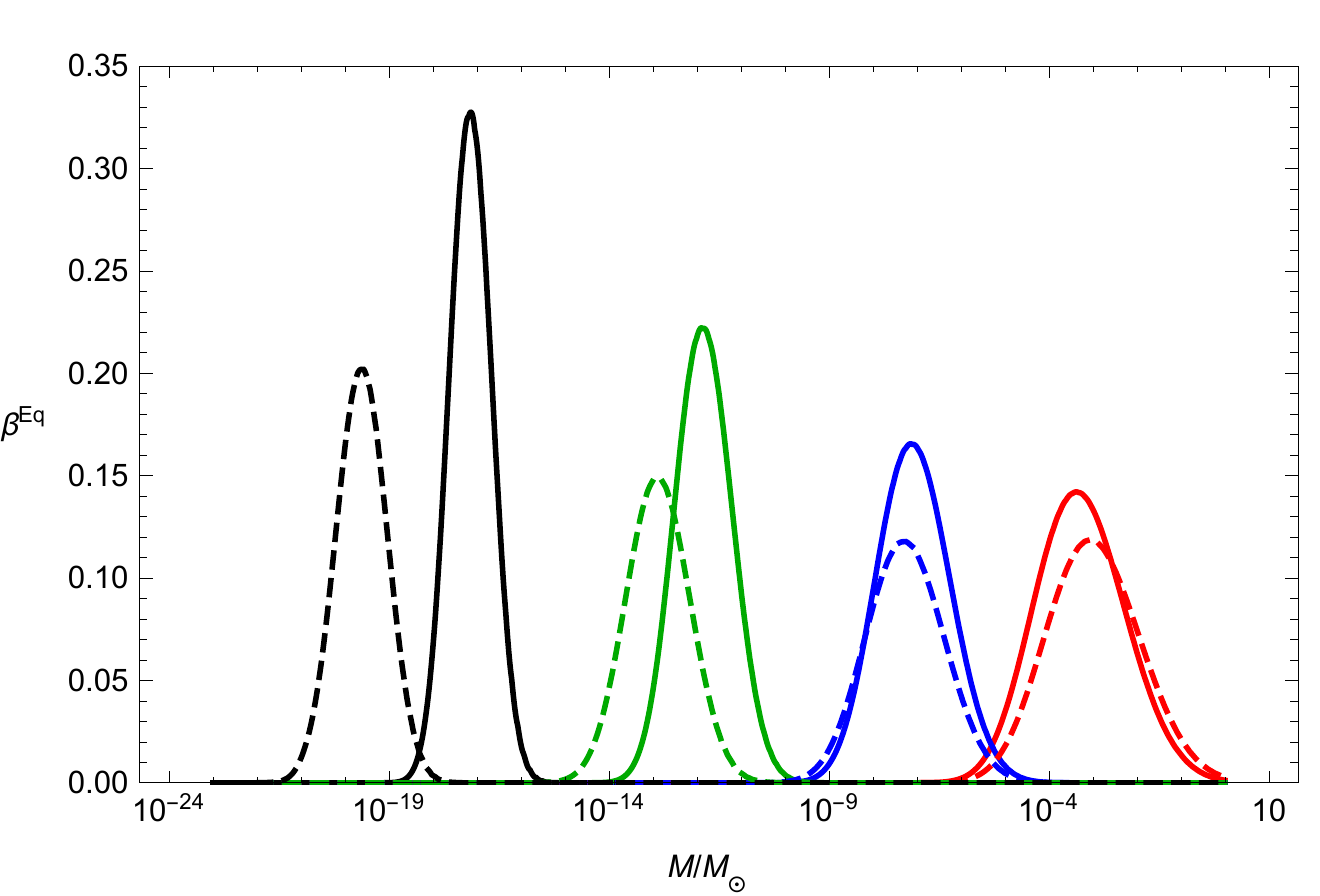}
	\caption{The relative energy density $\beta^{\rm Eq}$ for the hybrid-inflation case 
		with the variance $\sigma$ according to Eq.~\eqref{eq:sigma-hybrid}, 
		at the time of matter-radiation equality as a function of $M / M_{\odot}$.
		The solid curves assume standard black-hole production with horizon mass, 
		where the parameters are $\mu_{1} = 3 \times 10^5\.M_{\rm Pl}$, $M = 0.1\.M_{\rm Pl}$ for all graphs,
		and individually (right to left) 
		$\phi_{\crm} = 0.125\.M_{\rm Pl}$, $\delta_{\crm} = 1.43$ (red), 
		$\phi_{\crm} = 0.1\.M_{\rm Pl}$, $\delta_{\crm} = 1.54$ (blue), 
		$\phi_{\crm} = 0.1\.M_{\rm Pl}$, $\delta_{\crm} = 1.63$ (green), and
		$\phi_{\crm} = 0.075\.M_{\rm Pl}$, $\delta_{\crm} = 1.64$ (black). 
		The dashed curves have the same respective parameters, 
		but assume critical scaling according to Eq.~\eqref{eq:M-delta-scaling}, 
		using $\gamma = 0.36$.}
	\label{fig:beta-Eq-Hybrid-Comparison}
\end{figure}

Figure~\ref{fig:beta-Eq-Hybrid-Comparison} shows the ratio $\beta$ at the time of matter-radiation equality, where, again, the dashed lines depict the influence of critical collapse. Just as in the running mass case (Sec.~\ref{sec:Running--Mass-Inflation}), we see that the curves, appart from the right-most one, shift towards lower masses when critical collapse is accounted for. This originates from Eq.~\eqref{eq:Beta_normalisation} and becomes much more pronounced at lower masses. The relative overall amplitude in $\beta$ between horizon-mass collapse and critical collapse also decreases gradually with decreasing mass.

The parameters we have considered for the models depicted in Fig.~\ref{fig:beta-Eq-Hybrid-Comparison} are chosen to match those considered in \cite{Clesse:2015wea} to provide easy comparison with the literature. The authors of Ref.~\cite{Clesse:2015wea} used $\delta_{\crm}$ as a free parameter, which{\,---\,}given a set of model parameters{\,---\,}has been chosen such that all dark matter can consist of primordial black holes, yielding that all those values of $\delta_{\crm}$ are much higher than the numerically favoured $\delta_{\crm} \sim 0.45$.\footnote{Note that, contrary to the original paper of the investigated hybrid-inflation model \cite{Clesse:2015wea}, for the evaluation of $\beta$ we use the density contrast $\delta$ instead of the curvature perturbation $\zeta$ in order to establish comparability among the different models studied in this article. For a related discussion see Ref.~\cite{Young:2014ana}.} Of course, in a more proper treatment one would need to take this value and then look for implications on the model parameters. However, in this subsection we follow the mentioned literature for better comparison and focus on the implication of critical collapse. This also demonstrates that the critical collapse effects such as shift, broadening and rescaling is present for different values of $\delta_{\crm}$. From Fig.~\ref{fig:beta-Eq-Hybrid-Comparison} we can also see that the effects of the critical collapse are highly parameter dependent, and hence can not be treated by a constant rescaling or a constant shift if proper comparison to observations is to be made.

%% file: Section--Axion--like-Inflation.tex
A curvaton \cite{Lyth:2001nq, Lyth:2002my} is an extra field present during inflation. By decaying into standard model particles, its fluctuations can produce the curvature perturbations observed for instance in the CMB. These perturbations can of course also lead to production of primordial black holes.
		
In \cite{Kohri:2012yw,Kawasaki:2012wr} a curvaton model with an axion-like curvaton, i.e. a curvaton moving in an axion-like or natural inflation type potential, is described in order to produce primordial black holes, and this is the model that we will analyse further here. This model was originally described in \cite{Kasuya:2009up}, where the axion-like model is built into a supersymmetric framework, wherein the inflaton $\phi$ is the modulus, and the curvaton $\chi$ is related to the phase $\theta$ of a complex superfield $\Phi$. In practice the inflaton rolls down a potential of the form
\begin{align}
	V( \phi )
		&=
								\frac{ 1 }{ 2 }\.\lambda\,H^{2}\phi^{2}
								\; ,
\end{align}
where $H$ is the Hubble rate and $\lambda$ is a parameter of the theory which is derived from combinations of parameters in the supergravity theory. Because of its large mass, the inflaton rolls fast towards its minimum $\phi_{\rm min}$. Only after this time the curvaton is well-defined as $\chi = \phi_{\rm min} \theta \sim f \theta$ and becomes the primary degree of freedom of the superfield. The curvaton is assumed to move in an axion-like potential similar to that of natural inflation \cite{Freese:1990rb}
\begin{align}
	V_{\chi}
		&=
								\Lambda^{4}
								\left[
									1
									-
									\cos\!
									\left(
										\frac{ \chi }{ f }
									\right)
								\right]
		\simeq
								\frac{1}{2}\.m_{\chi}^{2}\.\chi^{2}
								\; ,
\end{align}
where the last equality holds when $\chi$ is close to $\chi_{\rm min} = 0$ and the curvaton mass is $m_{\chi} = \Lambda^{2} / f$. The particular shape of this potential, which is one preserving the shift symmetry peculiar to axions, is what makes this curvaton axion-like.
 
The power spectrum of primordial perturbations is generated by the combined perturbations from inflatons and curvatons,
\vs{-2mm}
\begin{align}
	\Pcal_{\zeta} (k)
		&=
								\Pcal_{\zeta,\mathrm{inf}}( k )
								+
								\Pcal_{\zeta,\mathrm{curv}}( k )
								\; .
\end{align}
The inflaton term is dominant on large scales (small $k$), and the second on small scales (large $k$). The inflaton perturbation is assumed to yield a near scale-invariant spectrum with $\Pcal_{\zeta,\mathrm{inf}}( k ) \simeq 2\times 10^{-9}$, in accordance with CMB observations such as WMAP \cite{Komatsu:2010fb} and Planck \cite{Planck:2013jfk, Ade:2015lrj}. This contribution should dominate up to at least $k \sim 1\,\mathrm{Mpc}^{-1}$. We define the crossing scale $k_{\crm}$ to be the scale at which curvaton and inflaton contributions to the power spectrum are equal. In addition, the scale $k_{\frm}$ is the scale at which the inflaton reaches its minimum $\phi_{\rm min} \sim f$ and the curvaton becomes well-defined. $M_{\crm}$ and $M_{\frm}$ are the horizon masses when these scales cross the horizon, respectively. Primordial black holes can not form before these horizon-crossing times, because the perturbations are too small when $M_{\mathrm{H}}> M_{\crm}$, and because no curvaton perturbations exist for $M_{\mathrm{H}} > M_{\frm}$.  $M_{\frm}$ can be found explicitly from the parameters of the theory and has the value
\begin{align}
	M_{\frm}
		&\approx
								10^{13 - 12 / ( n_{\chi} - 1 )}
								\left(
									\frac{g_{\frm}}{100}
								\right)^{-1/6}
								\left(
									\frac{ k_{\crm} }{\mathrm{Mpc}^{-1} }
								\right)^{-2}
								\left(
									\frac{ \Pcal_{\zeta, \mathrm{curv}}( k_{\frm} ) }{ 2 \times 10^{-3} }
								\right)^{- 2 / ( n_{\chi}- 1 )}
								M_{\odot}
								\; ,
\end{align}
where $g_{\frm}$ is the number of radiative effective degrees of freedom at the scale $k_{\frm}$. Throughout our consideration we will follow \cite{Kawasaki:2012wr} and assume $k_{\crm} = 1\,\mathrm{Mpc}^{-1}$ and $g_{\frm} = 100$.

Primordial black holes can not form from the density perturbations due to the inflaton, as these are constrained by the CMB observations. In contrast, when the curvaton power spectrum becomes dominant, it can have much larger power while still evading the bounds, producing large black holes. However the curvaton perturbations are assumed not to collapse to form primordial black holes before it has decayed to standard-model particles. Hence it can only form primordial black holes of a minimum mass $M_{\rm min}$ corresponding to black holes produced at its decay time and later. The exact value for the decay time and hence the minimum mass $M_{\rm min}$ is not known, but it should be smaller than the horizon mass at the time of big bang nucleosynthesis $10^{38}\,\grm$ in order not to interfere with this process and smaller than $M_{\frm}$ to yield any primordial black-hole production. In \cite{Kawasaki:2012wr} $M_{\rm min}/M_{\frm} = 10^{-8}, 10^{-3} $ is considered, hence we will do the same here.

It can be shown that the variance of the density power spectrum due to the curvaton perturbations in this model with an axion-like curvaton reads \cite{Kawasaki:2012wr}:
\begin{align}
	\sigma_\delta^{2}(M_{H})
		&=
								\frac{8}{81}\,\Pcal_{\zeta,\mathrm{curv}}( k_{\frm} )
								\left[
									\left(
										\frac{ M_{\frm} }{ M_{H} }
									\right)^{(n_{\chi}-1)/2}
									\gamma\mspace{-1mu}
									\left(
										\frac{ n_{\chi}-1 }{ 2 }, \frac{ M_{H} }{ M_{\frm} }
									\right)
									+
									E_{1}\!
									\left(
										\frac{ M_{H} }{ M_{H_{0}} }
									\right)
								\right]
								\label{eq:sigma-axion-like}
\end{align}
for horizon mass $M_{H} > M_{\rm min}$. For $M_{H}$ smaller than this value we assume the curvaton power spectrum which can transform into primordial black holes to be equal to zero. Due to inhomogeneities of the curvaton decay, this is not strictly true, however, as in \cite{Kawasaki:2012wr} we will take this to be a reasonable approximation. The curvaton spectral index 
\begin{align}
	n_{\chi} - 1
		&=
								3
								-
								3\.\sqrt{
										1
										-
										\frac{4}{9}\lambda
									\,}
\end{align}
is controlled by the parameter $\lambda$. By setting $\lambda \in (1,\.9 / 4]$ we can obtain a sufficiently blue power spectrum of curvature perturbations for the curvaton in order to produce primordial black holes at scales smaller than those constrained by the CMB. The minimum mass $M_{\rm min}$ which is defined by the decay time of the curvaton, protects the model from overproducing primordial black holes at very small scales in spite of the blue power spectrum. The functions $\gamma$ and $E_{1}$ are given by
\begin{subequations}
\begin{align}
	\gamma( a, x )
		&\equiv
								\int_{0}^{x}\d t\;t^{a - 1}\erm^{-t}
								\; ,
								\displaybreak[1]
								\\[2mm]
	E_{1}( x )
		&\equiv
								\int_{x}^{\infty}\d t\;\frac{ \erm^{-t} }{ t }
								\; ,
\end{align}
\end{subequations}
which are the lower incomplete gamma function and the exponential integral respectively. 

\begin{figure}
	\centering
	\includegraphics[scale=1,angle=0]{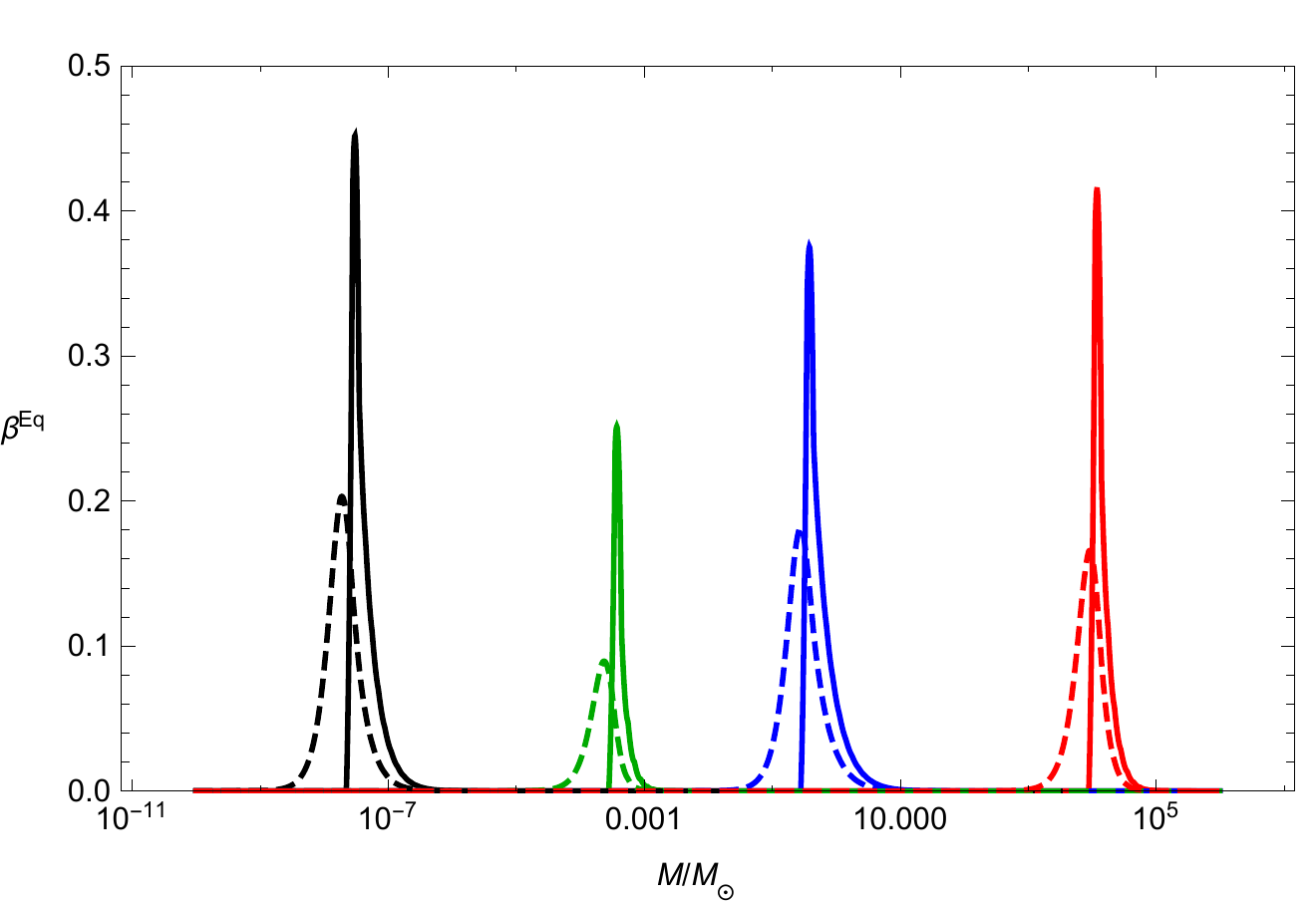}
	\caption{The relative energy density $\beta^{\rm Eq}$ for the axion-like curvaton inflation case 
		due to Eq.~\eqref{eq:sigma-axion-like}, at the time of radiation-matter equality as a function of $M / M_{\odot}$. 
		All curves have been produced with a primordial black hole density fluctuation threshold $\delta_{\crm} = 0.45$.
		The solid curves assume standard black-hole production with horizon mass, 
		where the parameters are (right to left) 
		$P_{\zeta}( k_{\frm} ) = 1.06\times 10^{-2}$, $M_{\rm min} = 10^{-3}M_{\frm}$, $\lambda = 2.1$ (red), 
		$P_{\zeta}( k_{\frm} ) = 2.92\times 10^{-3}$, $M_{\rm min} = 10^{-8}M_{\frm}$, $\lambda = 2.1$ (blue),
		$P_{\zeta}( k_{\frm} ) = 5.51\times 10^{-3}$, $M_{\rm min} = 10^{-3}M_{\frm}$, $\lambda = 1.2$ (green), and
		$P_{\zeta}( k_{\frm} ) = 1.93\times 10^{-3}$, $M_{\rm min} = 10^{-8}M_{\frm}$, $\lambda = 1.2$ (black). 
		The dashed curves have the same respective parameters, 
		but assume critical collapse according to Eq.~\eqref{eq:M-delta-scaling}.}
	\label{fig:beta-Eq-Axion}
\end{figure}

To illustrate the effects of critical collapse to the primordial black-hole production in this model we have chosen to consider models with the parameters $\lambda = 1.2$, $2.1$ and $M_{\rm min} / M_{\frm} = 10^{-8},10^{-3}$. The overall normalisation $\Pcal_{\zeta, \mathrm{curv}}( k_{\frm} )$ has been chosen in the range $10^{-3}$ -- $10^{-2}$ so as to produce a significant amount of black holes at the time of matter-radiation equality. This is done so that the primordial black holes can be the main contributor to dark matter today. The same is done in \cite{Kawasaki:2012wr} to show that the axion-like curvaton can produce primordial black holes to serve as dark matter. The results are shown in Fig.~\ref{fig:beta-Eq-Axion}. The solid lines are the results obtained when considering all the primordial black holes to be produced at horizon size. In this case all spectra produce a very sharp cutoff determined by the absolute value of $M_{\rm min}$ in each case. The subsequent decline is determined by the other parameters of the theory, but all cases that produce black holes that can make up the dominant part of the dark matter show quite narrow peaks. When the critical scaling Eq.~\eqref{eq:M-delta-scaling} is taken into account, the spectrum is widened, the peak is lowered and shifted towards lower values of the primordial black-hole mass.

%% file: Section--First--Order-Phase-Transitions.tex
Primordial black-hole production during first-order phase transitions were first suggested in \cite{Crawford:1982yz} and then revisited in \cite{Jedamzik:1996mr}. The amplicification of density perturbation due to the vanishing of the speed of sound during this transition was considered in detail for the QCD phase transition in \cite{Schmid:1998mx} and further developed to consider primordial black hole production specifically in \cite{Widerin:1998my}. A semi-analytic application to consider primordial black hole production in the QCD phase transition was made in \cite{Cardall:1998ne}. A numerical investigation of the critical collapse of these was made by Niemeyer and Jedamzik \cite{Jedamzik:1999am}, and shown to yield rather different results then what is found for primordial black hole production due to a peak in the power spectrum as was the case for the other models considered in this paper. 

Regardless of whether critical or horizon mass collapse is considered, the threshold value $\delta_{\crm}$ for collapse is significantly lowered during the first-order phase transition. In a regular radiation dominated scenario, the collapse of an overdensity to form a primordial black hole is counteracted by the radiation pressure of the fluid when the density $\delta < \delta_{\crm} \simeq 0.45$. During a first-order phase transition, however, two phases of fluids such as for instance quark-gluon plasma and hadrons coexist, and during this transition the expansion of the Universe can proceed at constant temperature by converting quark-gluon plasma to hadrons. Since the temperature is constant, the sound speed vanishes and hence the effective pressure that would otherwise slow down or prohibit the collapse of an overdensity is reduced. This then lowers the threshold value $\delta_{\crm}$ for collapse possibly as much as to $\delta_{\crm} \simeq 0.15-0.2$ \cite{Cardall:1998ne, Jedamzik:1999am}. These numbers were not obtained in the most updated numerical treatments of critical collapse, as these have not been applied to the first order phase transition scenario, though the lowering of the threshold should be generic. Therefore we choose to consider a range of different values of the threshold $\delta_{\crm}$ during the phase transition ranging from the usual background value $\delta_{\crm} = 0.45$ and down to the value $\delta_{\crm} = 0.2$ indicated in \cite{Cardall:1998ne, Jedamzik:1999am}. Furthermore, since the dynamics of the background changes, the scaling exponent $\gamma$ also changes with peak values above $\gamma \simeq 2$, depending on the dynamics of the transition (\cf~Ref.~\cite{Jedamzik:1999am}).

As the precise phase-transition dynamics is still unknown to a large degree, and since the main purpose of this work is to demonstrate the importance of critical scaling, we stick to a rather simple model for the changes of $\gamma \ra \tilde{\gamma}( M )$ and $\delta_{\crm} \ra \tilde{\delta}_{\crm}( M )$. We assume a first-order phase transition at the QCD energy scale {$T_{\rm QCD} \simeq 155$\,MeV \cite{Bhattacharya:2014ara}, which implies a horizon mass of $M_{\rm QCD} \sim 2\.M_{\odot} (T_{\rm QCD} / 100\,\text{MeV})^{-2} \approx M_{\odot}$ \cite{Jedamzik:1999am}. For its duration we will assume one Hubble time, and use the extremely simplified ad hoc scalings
\begin{subequations}
\begin{align}
	\tilde{\gamma}( M )
		&=
								\gamma
								+
								\left(
									\gamma_{\rm max}
									-
									\gamma
								\right)
								\exp
								\Big[
									-
									10\.
									\big(
										M / M_{\odot}
										-
										M_{\rm QCD} / M_{\odot}
									\big)^{2}
								\Big]
								\; ,
								\label{eq:gamma-tilde}
\end{align}
and
\begin{align}
	\tilde{\delta}_{\crm}( M )
		&=
								\delta_{\crm}
								\left(
									1
									-
									\Delta_{\crm}
								\right)
								\exp
								\Big[
									-
									10\.
									\big(
										M / M_{\odot}
										-
										M_{\rm QCD} / M_{\odot}
									\big)^{2}
								\Big]
								\; ,
								\label{eq:delta-c-tilde}
\end{align}
which is chosen in order to account for the results of \cite{Jedamzik:1999am} (\cf~Fig.~4 therein). However, we stress that, due to the large uncertainty on the form of $\tilde{\delta}_{\crm}( M )$, we regard the form of \eqref{eq:delta-c-tilde} as an approximate model to the actual change of the exponent $\gamma$ during the phase transition. Above, $\gamma_{\rm max}$ denotes the maximum value which $\tilde{\gamma}$ is assumed to take, and $\Delta_{\crm}$ is a constant shift which parametrises deviations with respect to its chosen base value of $\delta_{\crm} = 0.45$. As the physics of primordial black-hole formation through the QCD phase transition is not yet fully understood, we leave $\Delta_{\crm}$ as a free parameter which we vary in our subsequent study. Figure~\ref{fig:gamma-delta-Phase-Transition-Comparison} shows both of the above functions as a function of horizon mass in units of solar mass (\cf~figure caption for details on the parameters). We again stress that our aim is to demonstrate the effect of critical scaling, irrespective of whether the underlying model is the most realistic.

Though the collapse process is much more effective during the first-order QCD transition, the enhancement is not strong enough to produce primordial black holes from inflationary perturbations of a plain non-running red-tilted spectrum \cite{Schmid:1998mx, Widerin:1998my}. Therefore we use a running-mass model as described in Sec.~\ref{sec:Running--Mass-Inflation}, which would otherwise not produce primordial black holes, but which has an increase in the density power spectrum for the scales that cross horizon during the QCD phase transition.

In Fig.~\ref{fig:beta-Eq-Phase-Transition-Comparison-gammasset=2} we show the ratio $\beta$ at the time of matter-radiation equality, where we used an underlying running-mass inflation model which does not produce primordial black holes above $\beta \sim 10^{-6}$. The dashed lines depict the influence of critical collapse. Again we observe both a shift towards lower masses as well as a broadening with respect to the horizon-mass case. Due to the narrow-peaked nature of the black-hole production during the phase transition, plus the strong increase of $\gamma$ at this time, the effect of critical collapse is more pronounced then in any other of the three previously studied inflationary generated scenarios. As mentioned earlier the precise values for $\gamma$ and $\delta_{c}$ are not very well established as they were only computed with relatively limited computing power. If primordial black hole production from a first-order phase transition is to be studied in a more realistic scenario, better knowledge of the scaling and collapse behaviour during such a transition is required.

\end{subequations}
\begin{figure}
	\centering
	\includegraphics[scale=1,angle=0]{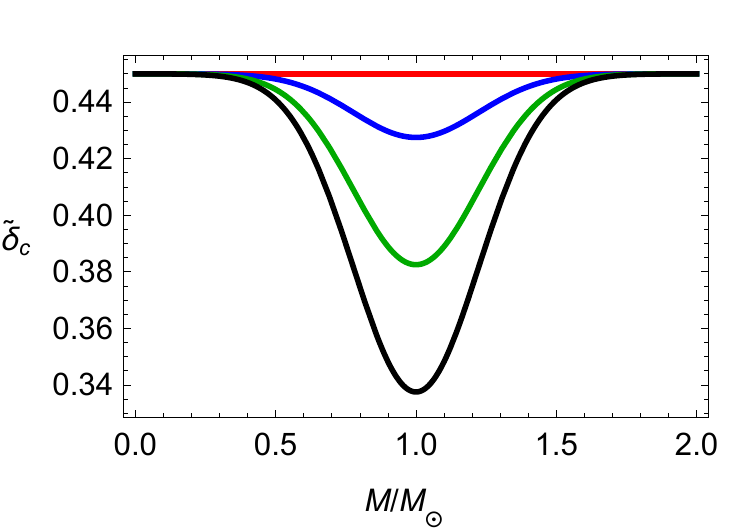}
	\qqq
	\includegraphics[scale=1,angle=0]{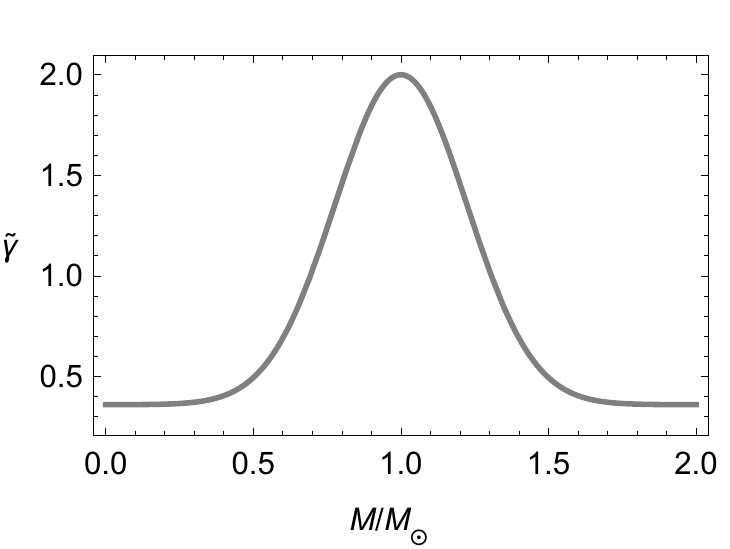}
	\caption{{\it Left panel:\.} Threshold $\tilde{\delta}_{\crm}$ [\cf~Eq.~\eqref{eq:delta-c-tilde}] as a function 
		of horizon mass over solar mass. The respective parameters are (top to bottom) 
		$\Delta_{\crm} = 0.45$ (red), 
		$\Delta_{\crm} = 0.4$ (blue), 
		$\Delta_{\crm} = 0.3$ (green), and
		$\Delta_{\crm} = 0.2$ (black).
		{\it Right panel:\.} Scaling exponent $\tilde{\gamma}$ [\cf~Eq.~\eqref{eq:gamma-tilde}] as a function 
		of horizon mass over solar mass for the choice of $\gamma_{\rm max} = 2$.}
	\label{fig:gamma-delta-Phase-Transition-Comparison}
\end{figure}

\begin{figure}
	\centering
	\includegraphics[scale=1,angle=0]{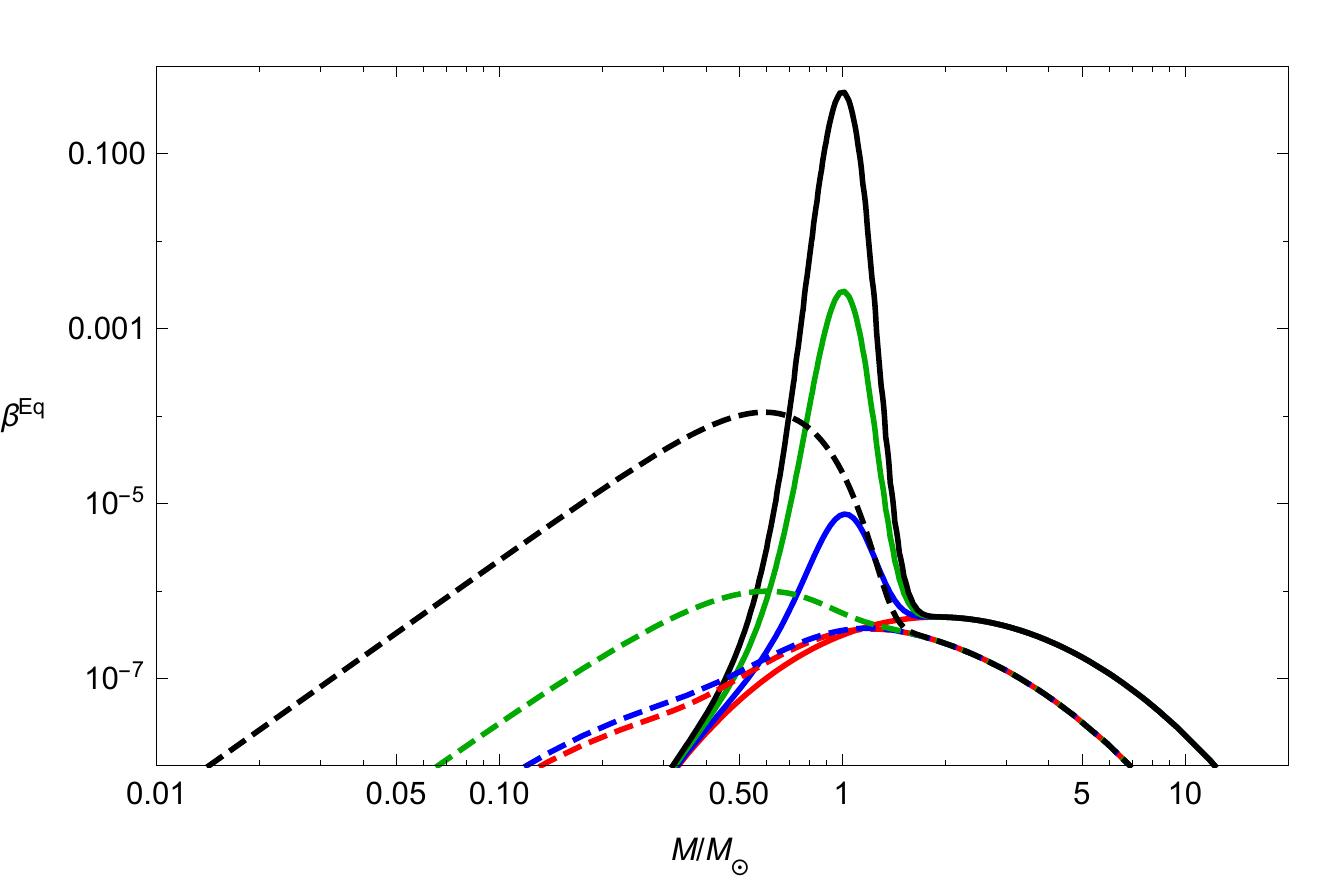}
	\caption{The relative energy density $\beta^{\rm Eq}$ for primordial black-hole production 
		during the QCD phase transition (assuming it is first order), 
		at the time of matter-radiation equality as a function of $M / M_{\odot}$.
		The inflationary model considered for the underlying power spectrum is that of the running mass inflation, 
		Eq.~\eqref{eq:sigma-running-mass}, with $a = 0.011$, $b = 0.02$, $c = - 0.0023645$, $\delta_{\crm} = 0.45$, 
		and $\gamma = 0.36$ (red, solid, lowermost).
		The solid curves assume standard black-hole production with horizon mass, 
		where the parameters are (bottom to top) 
		$\delta_{\rm c, max} = 0.4$ (blue), 
		$\delta_{\rm c, max} = 0.3$ (green), and
		$\delta_{\rm c, max} = 0.2$ (black).
		The dashed curves have the same respective parameters, 
		but assume critical scaling according to Eq.~\eqref{eq:M-delta-scaling},
		where $\gamma$ is given by Eq.~\eqref{eq:gamma-tilde} and we use $\gamma_{\rm max} = 2$.\vs{-2mm}}
	\label{fig:beta-Eq-Phase-Transition-Comparison-gammasset=2}
\end{figure}

%% file: Section--Summary-and-Outlook.tex
In this paper we have studied the formation of primordial black holes in several models of the early Universe. Specifically we have studied viable models which may produce primordial black holes in quantities comparable to the current dark matter abundance. 

In the literature such primordial black hole production have generally been approximated to yield black holes of horizon size at the time of formation. However, it was found and subsequently confirmed in several thorough numerical works that the primordial black holes are formed through critical collapse \cite{Choptuik:1992jv, Niemeyer:1998ac, Musco:2004ak, Musco:2008hv, Musco:2012au}. The critical collapse leads to a spectrum for the formation of primordial black holes at any time which follows the scaling law Eq.~\eqref{eq:scaling}, which leads to a peak of the mass spectrum at masses sometimes considerably smaller than the horizon mass.

Though the scaling law (in the context of primordial black holes) has been known for more than fifteen years, initial studies of its effects \cite{Yokoyama:1998xd, Green:1999xm} have been interpreted to mean that this effect is too small to have bearing and that the horizon mass approximation is still good.\footnote{To quote Ref.~\cite{Yokoyama:1998xd} precisely, the previous statement holds with the exception of primordial black holes with masses in the interval $4 \times 10^{14}\,\grm$ to $6 \times 10^{16}\,\grm$.} However, this study, which has gone through many of the prime candidates for viable primordial black-hole production which are not yet excluded by observations, show that this is not necessarily true anymore. The present constraints \cite{Carr:2009jm} on the mass distribution of primordial black holes, can be extremely tight for very specific mass ranges, but very loose for mass ranges very close to them. Hence a shift in mass distribution, lowering of peak value and widening of the distribution, as we have shown here for the various models in Figs.~\ref{fig:beta-Eq-Running-Mass-Comparison}, \ref{fig:beta-Eq-Hybrid-Comparison}, \ref{fig:beta-Eq-Axion}, and \ref{fig:beta-Eq-Phase-Transition-Comparison-gammasset=2} could mean the difference between exclusion or continued viability for a specific model. Even in mass regimes with less sharp boundaries between constrained and unconstrained mass ranges, we have shown that the changes to the mass distributions can be large, yielding alterations in the predictions of any such model. Since the shift and rescaling is also clearly model dependent, a constant shift to account for this as implemented in \cite{Carr:2009jm, Erfani:2015rqv} is reasonable when predicting constraints for the whole mass range, but not enough when considering the viability of any particular model.

The model parameters that we have chosen to consider here might not be entirely realistic, however, they are all more or less in accordance with models that we found in considerations done in the literature. Apart from (p)reheating models, we have also covered what we consider to be a representative set of models for primordial black-hole production. For future work, treatment of both more realistic parameters and (p)reheating models is of course possible. A better understanding of the collapse process and parameters during first-order phase transitions would also be important for a full treatment of these phenomena. Regardless of this we consider our results as fairly general and therefore suggest that the critical-collapse scaling should be taken into account when considering primordial black hole production in the future.

In addition to this main result, we note that when primordial black holes from running-mass inflation, the inclusion of first- and second-order running is not sufficient to yield viable production, as this leads to potentially unfeasibly large production of very light primordial black holes as can be seen from Fig.~\ref{fig:beta-form-without-running^3}. These are highly constrained by observations and hence theories that go only to this order are automatically ruled out. When the third-order running is included, this pathology can be avoided as shown in Fig.~\ref{fig:beta-Eq-Running-Mass-Comparison}. This was not considered in the original treatment \cite{Kawasaki:2012wr} where the focus was only on the production of sufficient amounts of primordial black holes at a certain high/intermediate mass. However, such a third-order running constitutes one additional term in the expansion of the power spectrum, and hence noticing the mentioned pathologies brings us one step closer to unveiling the true inflationary dynamics.